\documentclass[aps,pre,noshowpacs,twocolumn]{revtex4}
\usepackage{bm}
\usepackage{graphicx}
\usepackage{mathtools}
\usepackage{color, soul}
\usepackage{changes}

\usepackage{soul}
\usepackage{hyperref}
\usepackage[11pt]{moresize}

\begin{document}
\title{The size of the immune repertoire of bacteria}

\author{Serena Bradde}
\affiliation{American Physical Society, Ridge, NY,  \& David Rittenhouse Laboratories, University of Pennsylvania, Philadelphia, PA 19104}
\author{Armita Nourmohammad}
\affiliation{Max Planck Institute for Dynamics and Self-organization, Am Fa\ss berg 17, 37077 G\"ottingen, Germany\\
Department of Physics, University of Washington, 3910 15th Ave Northeast, Seattle, WA 98195}
\author{Sid{hartha} Goyal}
\affiliation{Department of Physics, University of Toronto, Toronto, ON M5S 1A7}
\author{Vijay Balasubramanian}
\affiliation{David Rittenhouse Laboratories, University of Pennsylvania, Philadelphia, PA 19104}
\begin{abstract}
Some bacteria and archaea possess an immune system, based on the CRISPR-Cas mechanism, that confers adaptive immunity against phage.  In such species, individual bacteria maintain a ``cassette'' of viral DNA elements called spacers as a memory of past infections.   The typical cassette contains a few dozen spacers.  Given that bacteria can have very large genomes, and since having more spacers should confer a better memory,  it is puzzling that so little genetic space would be devoted by bacteria to their adaptive immune system. Here, we identify  a fundamental trade-off between the  size of the bacterial immune repertoire and effectiveness of response to a given threat, and show how this tradeoff imposes a limit on the optimal size of the CRISPR cassette. 
\end{abstract}

\maketitle

\section{Introduction}

All living things from bacteria to  whales are under constant threat from viruses.  To defend against these threats many organisms have developed innate mechanisms that make it harder for infections to occur (e.g., by impeding entry of viruses through the cell membrane), or that respond universally to the presence of infections (e.g. through inflammatory responses)~\cite{Labrie2010}.   While such innate defenses are very important, they likely cannot be as effective as defenses that specifically target particular infections.  The challenge for developing specific responses is that they require a mechanism for {\it learning} to identify each incident virus, and a mechanism for {\it remembering} the {defended targets}.  Vertebrates implement such a learning-and-memory approach in their adaptive immune system, which produces novel antibodies through random genomic recombination, selects effective immune elements when they bind to invaders, and then maintains a memory pool to guard against future invasions.   Recently, we have learned that some bacteria also enjoy adaptive immunity \cite{Barrangou2014} -- they maintain a cassette of ``spacers'' (snippets of DNA from previously encountered phage), and use these spacers via the CRISPR-Cas mechanism to identify and clear recurring infections by similar viruses.

The CRISPR-Cas mechanism for immunity has three stages \cite{Barrangou2014,Bonomo2018}.  Following a first encounter with a virus, after a successful defense through another mechanism or if the virus is ineffective for some reason \cite{Hynes:2014aa}, some of the { Cas} proteins recruit pieces of viral DNA and integrate these ``spacers'' into an array separated by palindromic repeated sequences in the CRIPSR locus of the bacterial genome.   This array defines the CRISPR ``cassette'' which represents the memory of past infections.  In the second stage, the entire CRIPSR locus is transcribed into RNA through a mechanism that depends on the CRISPR type of the bacterium.   Finally, invading sequences are recognized by base-pairing with the CRISPR RNA.  A successful match triggers cleavage of the virus through the action of complexes formed by other  Cas proteins.  

 In lab experiments that expose naive bacteria, which start without spacers, to carefully controlled environments,  the CRISPR cassette is often small, consisting of a few spacers acquired during interactions with phage.  Wild-type bacteria can have larger cassettes, but even these contain only a few tens  to at most a few hundreds  of spacers \cite{Devashish_2015,Martynov2017,grissa2007crisprdb, Barrangou2014}.  Metagenomic analysis of the human gut microbiome has revealed CRISPR cassettes with an average size of twelve spacers \cite{Mangericao2016}. A broader analysis of all sequenced bacteria and archeae found cassette lengths clustered in the twenty to forty range~\cite{Grissa:2007}. Similarly, a study of 124 strains of {\it Streptococcus thermophilus} revealed an average cassette size of 33~\cite{Horvath:2008}.
 
These findings lead to a puzzle.   Why do bacteria maintain such small memories in their adaptive immune systems given that they have probably been exposed over generations to thousands of species of phage?  Certainly, the size of the genome is not a constraint since bacterial genome sizes lie in the range of millions of base pairs. One idea could be that viruses evolve so quickly or are encountered so rarely that a deep memory is not useful \cite{Martynov2017}.  Here we propose an alternative hypothesis:  given limited resources for Cas-complex formation \cite{Vale2015a},  a deep memory imposes an opportunity cost by reducing the chance for the spacer specific to an invader to be activated in time to cleave the virus before it reproduces.  Thus, there is a tradeoff between effectiveness of the defense and the depth of memory. 
 
 We analyze this tradeoff quantitatively and demonstrate that it predicts an optimal size for the CRISPR cassette.   This optimum  is controlled by the number of {Cas} complexes that are available for cleaving viruses, and by the diversity of the phage landscape.   In the limit that the phage landscape is very diverse, the size of the cassette is largely limited by the number of {Cas} proteins that the bacterium can produce while carrying on its other functions  \cite{Vale2015a}. {Using the biologically relevant range for {Cas} protein concentration in a bacterium, we show that the optimal cassette size should typically lie in the range of 10-100 spacers, consistent with genomic observations \cite{Devashish_2015,Martynov2017, Mangericao2016,Grissa:2007,Horvath:2008}.

\begin{figure}[t]
\includegraphics[width=.5\textwidth]{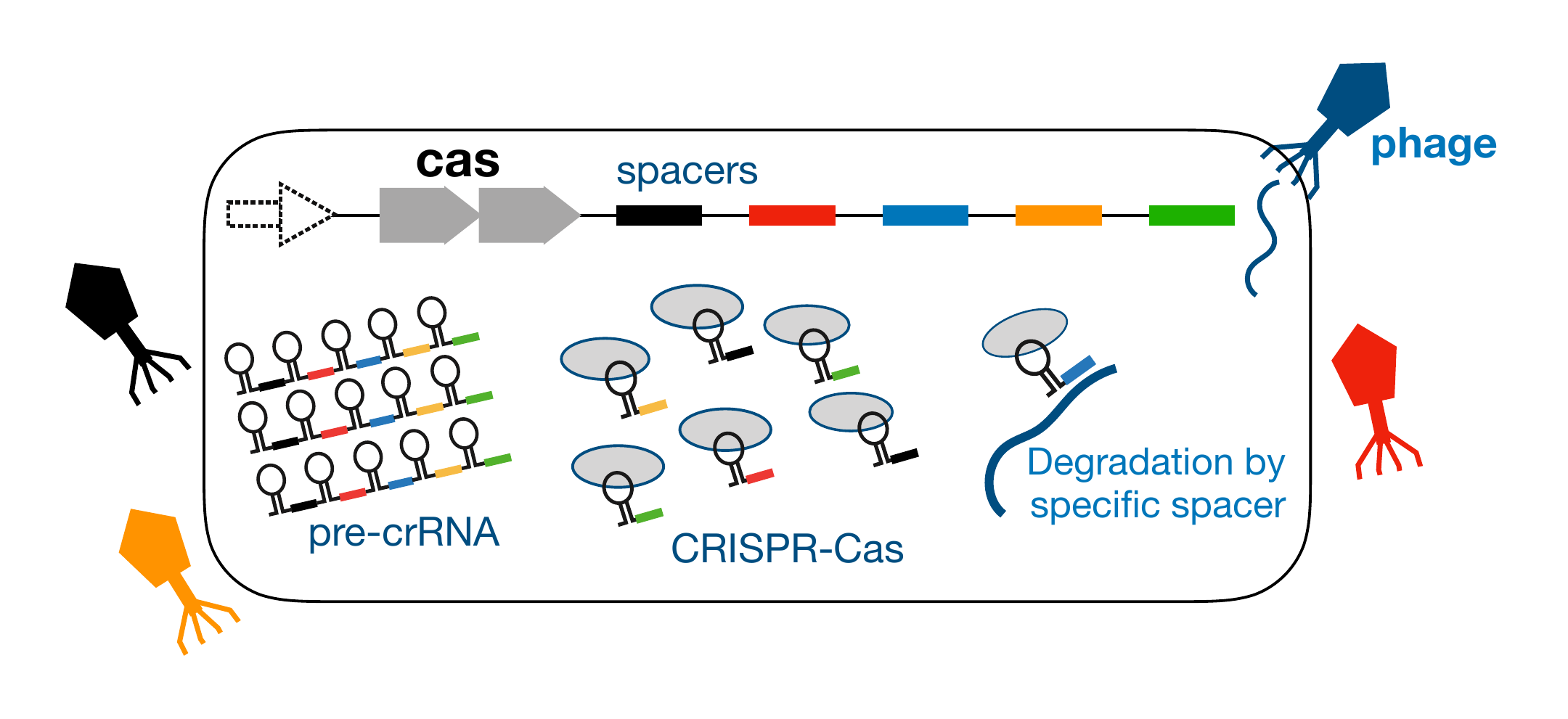}
\caption{\small{\bf CRISPR immunity in bacteria.}  A bacterium (bordered rectangle) with CRISPR machinery encounters a diverse set of phages (colors). The  CRISPR-Cas locus is transcribed and then processed to bind {Cas} proteins (gray ovals) with distinct spacers (colors), thus producing CRISPR-Cas complexes. The complex with a spacer that is specific to the injected phage DNA (same color) can degrade the viral material and protect the bacterium from infection. 
\label{fig:schematic}
}
\end{figure}

\section{Results}

\subsection{CRISPR as a probabilistic memory of phage}
We consider a model of infection where bacteria encounter $j=1,\ldots,K$ types of phage, each with probability $f_j$.   For simplicity, all types of phage are taken to be equally infectious and to have similar growth rates, conditions that are easily  relaxed.  In the CRISPR mechanism for adaptive immunity, bacteria incorporate snippets of phage DNA (``spacers") into the CRISPR cassette. Upon later infection, the bacteria recruit CRISPR-Cas complexes with  spacers that match the invading phage to cleave the viral DNA (Fig.~\ref{fig:schematic}).

Suppose that the cassette contains $L$ spacers, and that an individual bacterium maintains a population of $N_p$  { Cas} protein complexes that can be recruited to cleave invaders.   The cassette configuration can be characterized in terms of  a vector ${\bf s} = \{s_1\cdots s_K\}$ with entries counting the number of spacers specific to each phage type.    The cassette size is the sum of $s_j$, i.e. $\sum_j s_j = L$, and quantifies the amount of immune memory stored by an individual bacterium.  We describe the phage configuration in a given infection event as a vector ${\bf v}$ of length $K$ with entries indicating presence (1) or absence (0) of each viral type.  Finally, we define  the configuration of complexes ${\bf d} = \{d_1\cdots d_K\}$ as a vector with entries counting the number complexes specific to each phage during the CRISPR response.  The total number of complexes is the sum of $d_j$, i.e. $\sum_j d_j = N_p$.     In terms of these variables,  the probability of surviving a phage infection  using the CRISPR-Cas defense mechanism is
\begin{equation}\label{eqn:probid}
\begin{aligned}
P_\text{survival} =  1-\Big(
&\sum_{{\bf v}}
p_V({\bf v}) \sum_{s_1 + s_2 + \cdots = L} p_S( {\bf s} | L)  \\ &  
\times \sum_{ d_1 + d_2 + \cdots = N_p}  \big[1- \alpha ({\bf v}, {\bf d} )\big]  q( {\bf d} | {\bf s})\Big) \, .
\end{aligned}
\end{equation}
Here,  $p_V({\bf v})$ is the probability of encountering the phage configuration ${\bf v}$,  $p_S( {\bf s}  |L )$ is the  probability of having  a cassette configuration ${\bf s}$ of length $L$, $\alpha({\bf v}, {\bf d})$ is the probability of detecting all the viral types present in ${\bf v}$ given the configuration ${\bf d}$ of complexes,  and $q({\bf d} | {\bf s})$ is the probability of producing the  {CRISPR-Cas configuration}  ${\bf d}$   given the set of spacers ${\bf s}$ {and $N_p$ Cas protein complexes}.  

The form of the detection probability function $\alpha ({\bf v}, {\bf d})$  depends on the specific mechanism used by the CRISPR machinery to bind and degrade a phage. 
However, 
a critical number of specific complexes, $d_c$, is required   for the CRISPR machinery to achieve targeting at the speeds measured in experiments \cite{Jones1420}.   Thus, we consider two  functional forms for the probability that a particular phage type can be detected with $d$ specific complexes: {(i) a hard constraint $\alpha(d)=\theta(d-d_c)$, where $\theta(x)$ is step function and is 0 if $x<0$ and 1 if $x>0$, and (ii) a soft constraint $\alpha(d) = {d^h \over d^h + d_c^h}$.}  In both cases $d_c$ is an efficacy parameter that depends on biochemical rates, and  determines the threshold on $d$ below which detection is rare and above which detection is common.  The first functional form (step function) describes  switch-like behavior where complexes bind to phage DNA if they exceed a certain concentration $d>d_c$.  The second form mimics a Hill-like response, where the chance of binding increases gradually with the number of complexes.  Here, we allow the binding of Cas-complexes with phage DNA binding to be cooperative.  As the cooperativity $h$ increases, the binding  behavior becomes increasingly switch-like.

\begin{figure}
\includegraphics[width=.4\textwidth,angle=90]{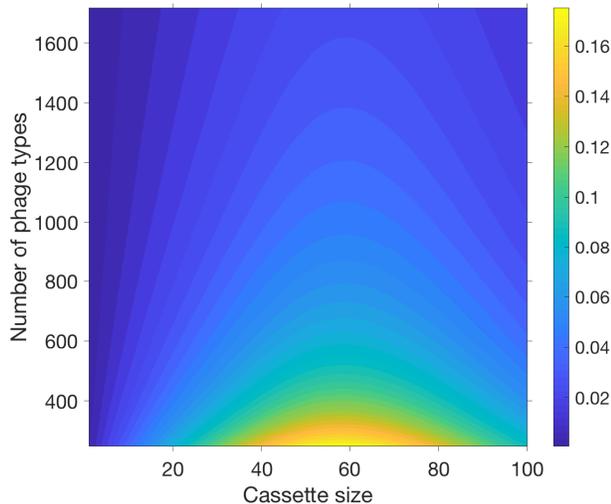}
\caption{\small {\bf There is an optimal amount of immune memory.} The  heatmap shows the probability of surviving a phage infection, $P_{\text{survival}}$, as a function of the cassette size and phage diversity with $N_p=700$ Cas complexes.   $P_{\text{survival}}$ can be interpreted as the fractional population size that will persist after sequential phage attacks if CRISPR is the only defense mechanism.  The detection probability function is a step function $\alpha(d)=\theta(d-d_c)$ with threshold $d_c=8$, implying that detection of  a phage requires at least $d_c$ complexes bound to  the corresponding spacers.   For any number of phage types there is an optimal cassette size.  See Fig.~S1 for different choices of $N_p$, $d_c$, and functional forms for $\alpha$.}
\label{fig:optimalLength}
\end{figure}

\subsection{There is an optimal amount of memory}

In realistic settings we can make simplifying assumptions about the general model in eq.~\ref{eqn:probid}.  For example, we can assume that  successful infections of a bacterium  by different phages occur with low probability and are independent.  Of course, a  given bacterium can be infected by multiple phages over its lifetime.  Since the probability that a bacterium simultaneously encounters multiple phages is small, we  assume that  encounters are sequential (i.e., the  viral configuration vector ${\bf v}$ has a single nonzero entry).  
 
Second, we  assume that  a bacterium's lineage encounters many and diverse phage types over multiple generations, i.e., $K \gg 1$.   Because phages mutate readily, there is  subtlety about what defines a ``type''.   
We use a functional definition --  a ``type'' of phage is defined by its specific recognition by a given spacer.  Sometimes, after a bacterium becomes immune to a phage, single point mutations in the virus can produces ``escapers'' that evade recognition.  By our definition these escapers are effectively a new type of virus that the bacterial population must deal with sequentially in future infections \cite{Deveau2008,Heidelberg2009, Sun2013,Houte2016}.

\begin{figure*}[t]
\includegraphics[width=.28\textwidth]{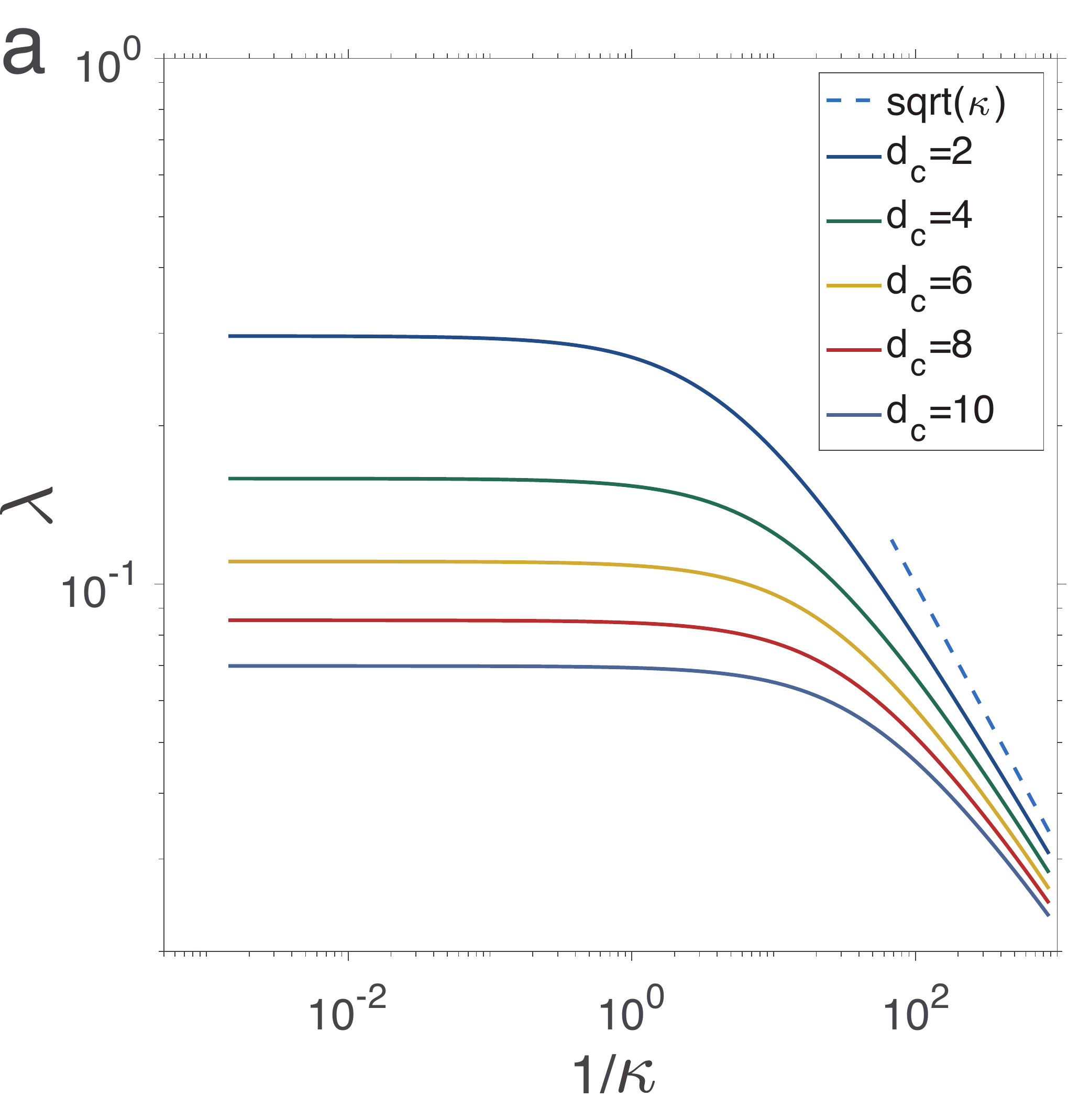}
\includegraphics[width=.28\textwidth]{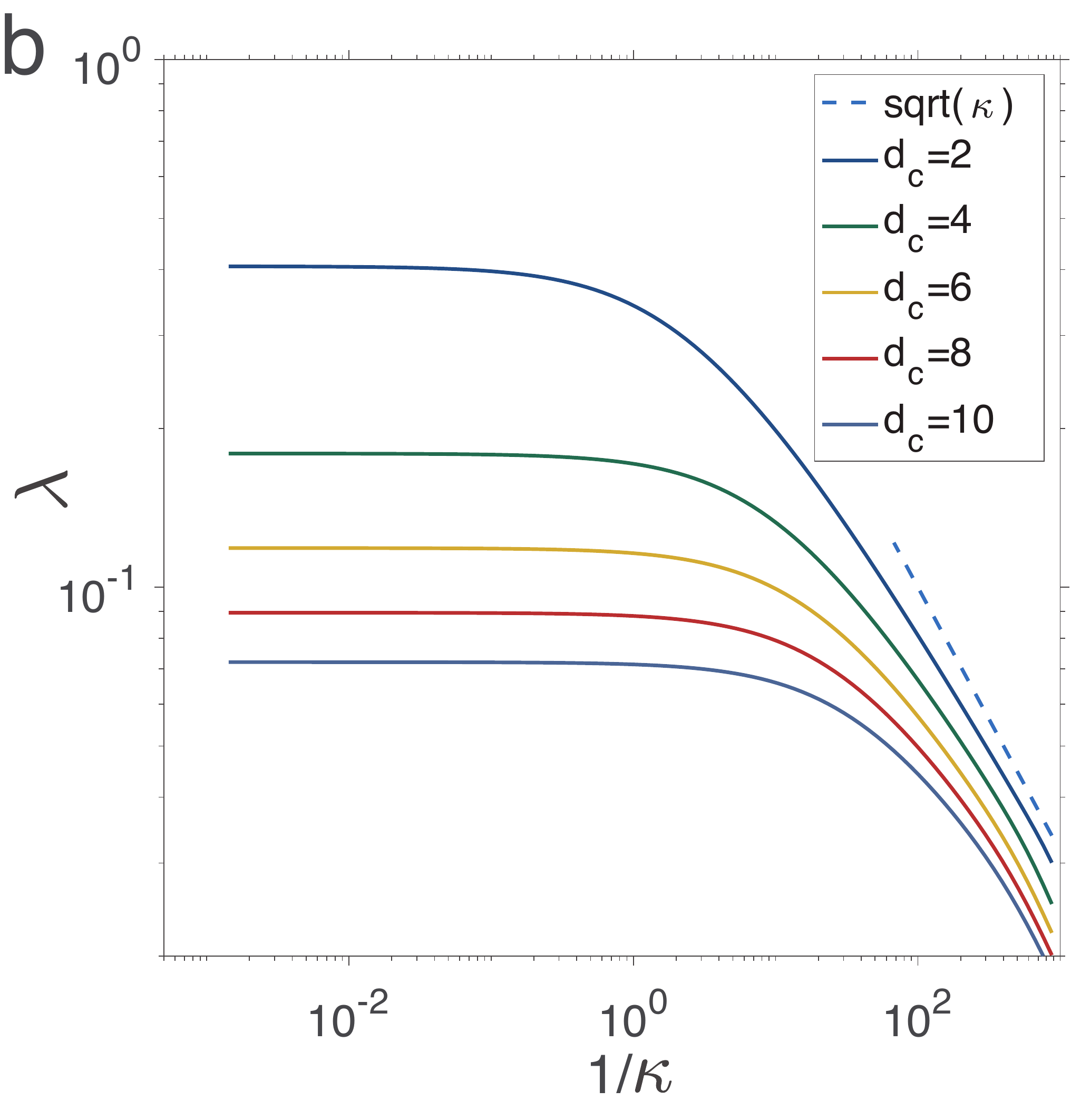}
\includegraphics[width=.28\textwidth]{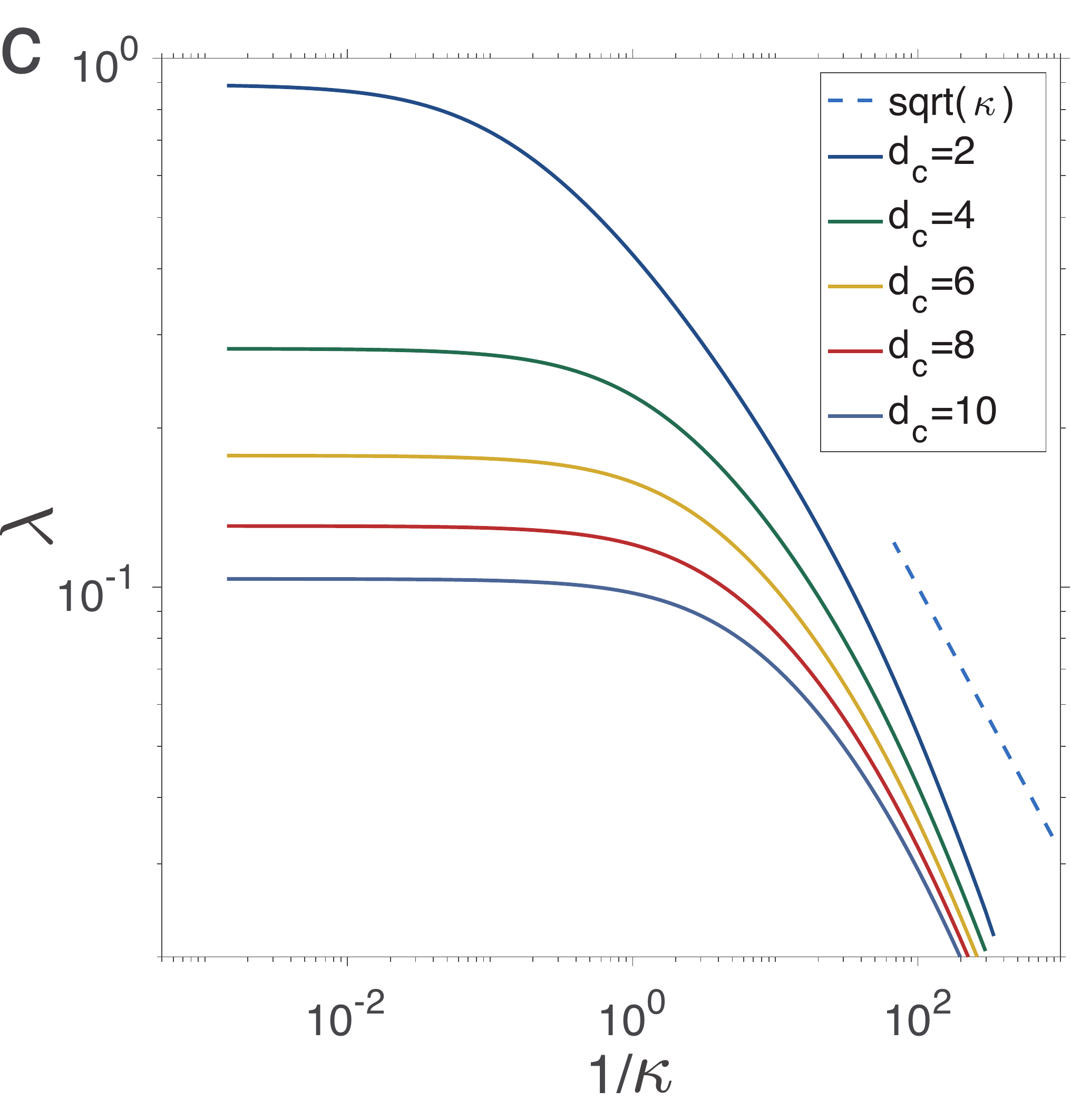}
\caption{{\bf Optimal amount of immune memory depends on  viral diversity.} The panels show the optimal cassette size ($L$) relative to the number of Cas complexes  ($N_p$) parameterized as $\lambda = L/N_p$, as a function of the viral diversity ($K$) relative to the number of complexes parameterized as $\kappa = K/N_p$.  We examine CRISPR machineries with different detection probability functions: (a) switch-like detection probability with a step function $\alpha(d)=\theta(d-d_c)$, and a smoother model $\alpha(d) = d^h/ (d^h + d_c^h)$ with (b) $h = 10$ leading to nearly switch-like detection probability, and  (c)  $h = 2$ leading to a softer transition between low and high detection probability.  Here $d_c$ is an effective threshold on the number of complexes ($d$) required for detecting phage with high probability.    }
\label{fig:numericald2}
\end{figure*}
 
 {Third,} we assume that spacers are uniformly sampled from the phage distribution over time.  This distribution is not known experimentally, so we  make a minimal assumption that all phage types are equally likely and occur with probability $1/K$.  This is a conservative assumption, because bacterial immune memory confers the least advantage when faced with an unbiased (i.e., a minimally informative) phage environment.
In effect, we are not interested here in the short-time co-evolutionary arms race between bacterium and phage, and are instead considering the long-term {statistical features}.

Finally, we assume that phage encounters from which spacers are acquired occur randomly.  Thus each spacer in the CRISPR cassette has a probability $1/K$ of being specific for a given phage.
  Since the  cassette size is  much smaller than the number of viral types ($L \ll K$), it follows that the cassette will either have one or no spacer that can target  a particular phage type -- in other words $s_i = 0,1$.     We also assume that all spacers along the cassette  are equally expressed, in the absence of alternative experimental evidence.   A biased expression of spacers may explain finite cassette sizes \cite{Martynov2017}, but our model does not require such additional assumptions.

We derive an expression  for the probability to survive a phage infection, given the cassette size $L$, the number of { Cas} complexes $N_p$ and the diversity of the phage population $K$ (Methods, Eq.~\ref{eqn:probfinal2}).  Across a wide range of parameters and for various choices of  detection probability functions {$\alpha(\bf{v}, \bf{d})$}, we find that there is an optimal amount of memory, consisting of a few tens of spacers in the CRISPR cassette, to maximize survival probability (Fig.~\ref{fig:optimalLength} and  Fig.~S1).  
The optimum occurs because there is a trade-off between the amount of  stored memory in CRISPR cassettes and  the efficacy with which a bacterium can utilize its limited resources (i.e., { Cas} proteins) to turn memory into a functional response. If the CRISPR cassette is too small,  bacteria do remember past phage encounters well enough to defend against future infections.  On the other hand,  if the cassette is too large, Cas complexes  bind too infrequently to the correct spacer to provide  effective immunity against a particular invading virus. The optimal amount of immune memory (cassette size) should lie in between these two extremes, with details that depend on the phage diversity, the number of {{ Cas}} complexes, and the detection probability function {$\alpha(\bf{v}, \bf{d})$} {(Fig.~\ref{fig:optimalLength} and Methods).}

\subsection{Optimal memory depends on phage diversity}
How does the optimal amount of CRISPR memory depend on the diversity of viral threats?   If there are relatively few types of phage,  an optimal strategy for a bacterium would be  to maintain an effective memory for most threats and to match the viral variants with  a cassette whose size grows with viral diversity. This matching strategy will eventually fail as the diversity of the phage population increases if  the number of {{Cas}} proteins ($N_p$) is limited. We examined this tradeoff by measuring the optimal cassette size as we varied the viral diversity ($K$) while  keeping the CRISPR machinery ($N_p$ and the detection probability $\alpha(\bf{v}, \bf{d})$) fixed. 

To characterize viral diversity, we defined a parameter $\kappa =K/N_p$ as the ratio of the number of phage types ($K$) and the number of Cas complexes ($N_p$). When phage diversity is low ($\kappa\leq1$) the optimal amount of memory (cassette size) increases sub-linearly with viral heterogeneity (Fig.~\ref{fig:numericald2}), approximately as a power law. When the detection probability  $\alpha(\bf{v}, \bf{d})$ is nearly switch-like,  the optimal cassette size scales approximately as $L \sim \sqrt{K}$ (Fig.~\ref{fig:numericald2}A,B).   This implies that when viral diversity is low the amount of memory should increases with the diversity,  but it is actually beneficial not to retain a memory of all prior phage encounters.   Forgetting some encounters will allow the bacterium to mount a stronger response against future threats by engaging a larger number of {Cas} complexes for the threats that are remembered.   This sub-linearity in the optimal amount of memory becomes stronger as the  the number of phage-specific CRISPR-{Cas} complexes necessary for an effective response, $d_c$, increases.

When phage diversity is high ($\kappa> 1$), the optimal amount of memory depends on  the CRISPR mechanism via the response threshold $d_c$, but is  independent of viral heterogeneity so long as $d_c \geq 2$ (Fig.~\ref{fig:numericald2}; see SI for discussion of the special case $d_c = 1$). In nature, phage are expected to be very diverse ($K \gg N_p$).  Thus, our model predicts  that the cassette size of a bacterium is determined by the expression level of {{Cas}} complexes $N_p$  and the detection threshold $d_c$ of the particular CRISPR mechanism that is used by the species.
\begin{figure*}
\includegraphics[width=.3\textwidth]{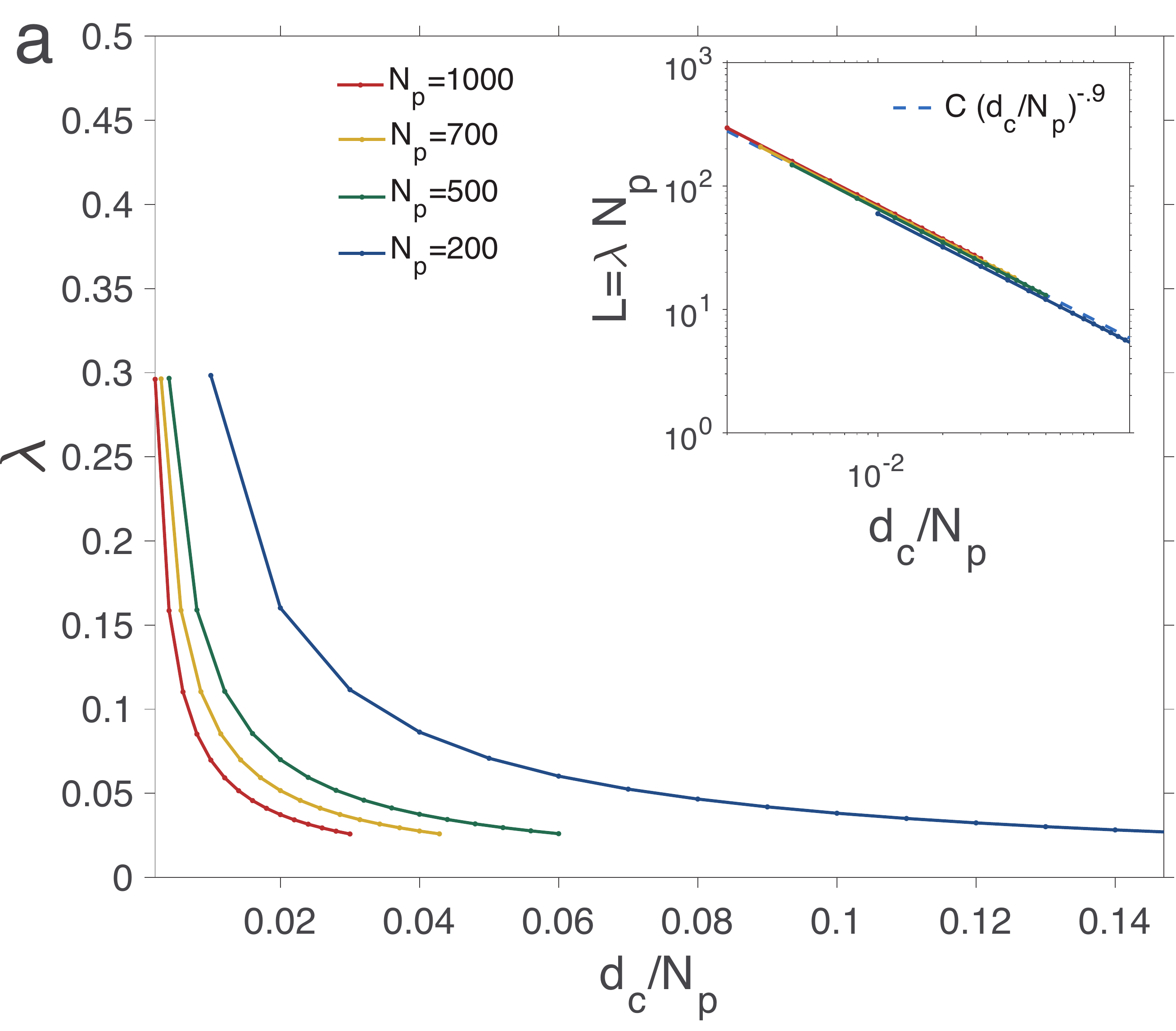}
\includegraphics[width=.3\textwidth]{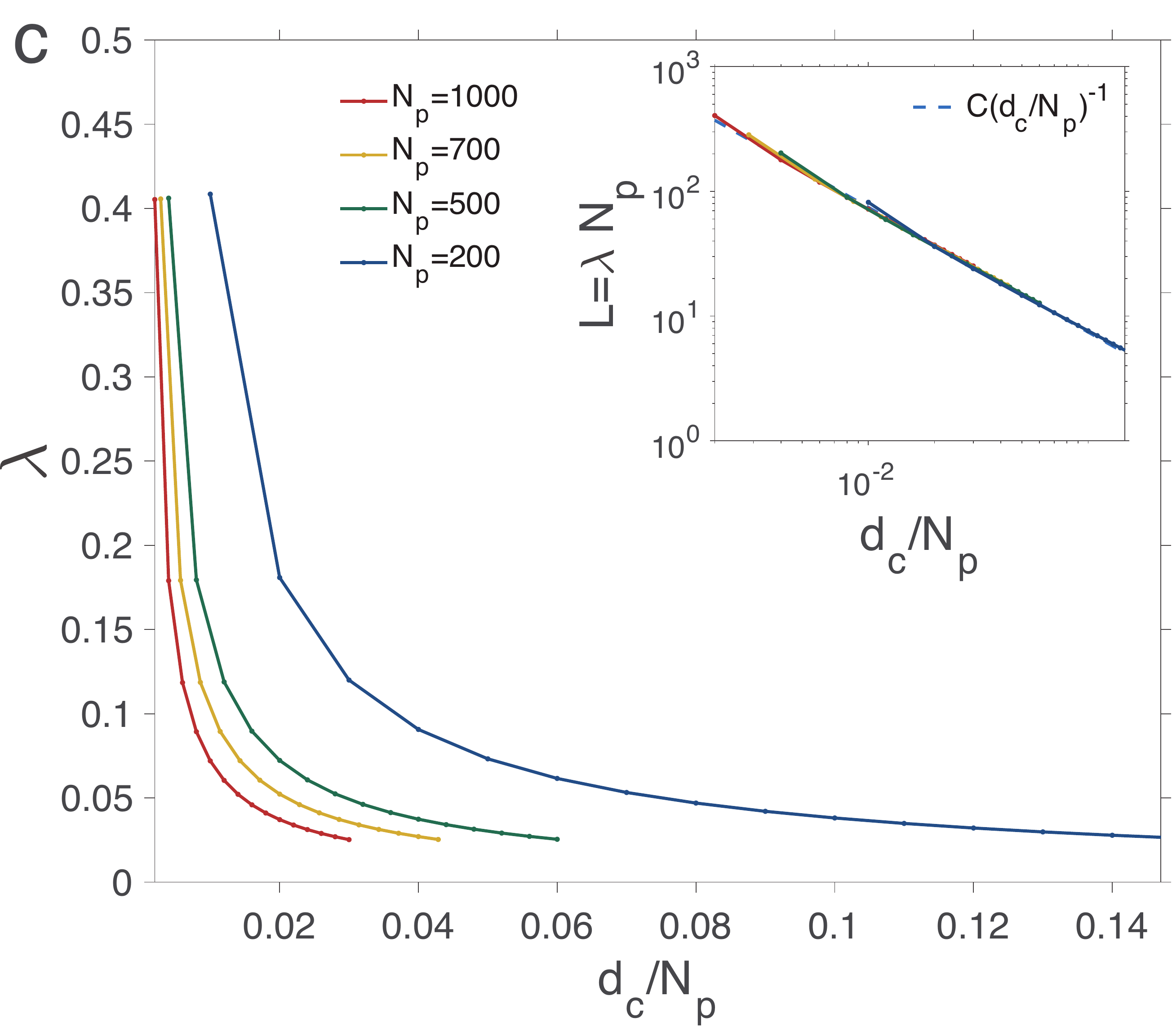}
\includegraphics[width=.3\textwidth]{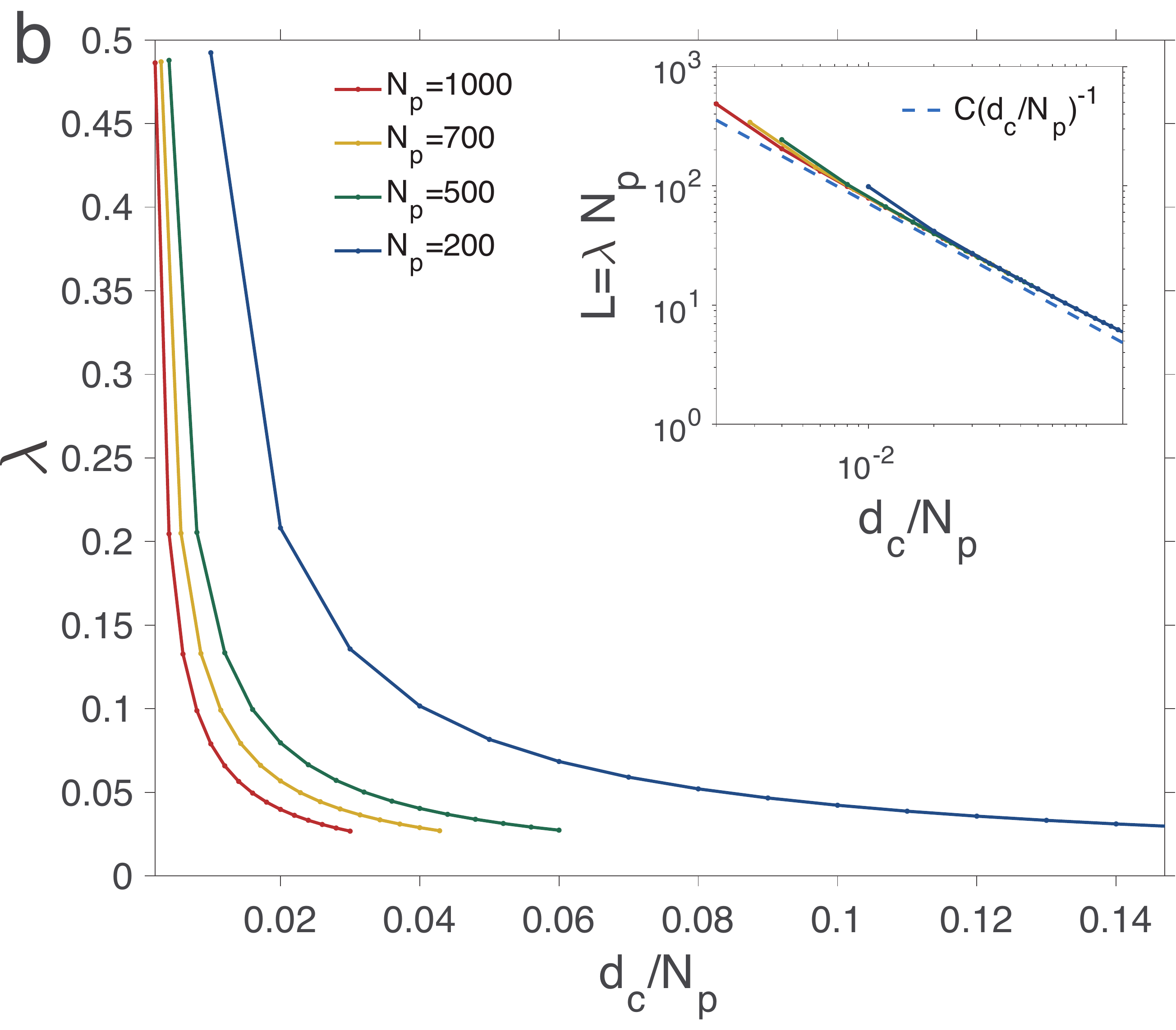}
\caption{\small {\bf Optimal amount of immune memory  depends on the detection threshold.}
The figure shows the optimal cassette size relative to the number of {Cas} proteins, $\lambda= L/N_p$, as a function of the threshold to detect a phage, also relative to the number of Cas proteins, $d_c/N_p$.  We consider different functional forms of the detection probability: (a) Step function with a sharp detection threshold   $\alpha(d)=\theta(d-d_c)$ and (b-c) Hill functions, $\alpha(d) = d^h/ (d^h + d_c^h)$, with $h=10$ in (b) and $h=3$ in (c). Phage types are taken to be 1000-fold more numerous than the number of complexes ($\kappa = K/N_p = 1000$).   Inset: Optimal cassette size scales as $L = C (d_c/N_p)^{-\beta}$  over a realistic range of values for the number of complexes and phage detection thresholds in a single bacterial cell.   Best fits are shown in each case with (a) $\beta=0.9$ and $C \sim 1$,  (b) $\beta=1$, $C\simeq 0.7$, and (c) $\beta=1$, $C \simeq 0.8$.
\label{fig:numericald3}}
\end{figure*}

\subsection{Memory  increases with detection efficacy}
Detection  efficacy depends on two key parameters: (i) the detection threshold $d_c$ and (ii) the number of available {Cas} proteins $N_p$. In Fig.~\ref{fig:numericald3}, we show that the optimal cassette size for defending against  a diverse phage population ($\kappa = K/N_p \gg 1$) decays as a power law of the detection threshold $\sim (d_c/N_p)^{-\beta}$ with an exponent $\beta \simeq 1 $ (Methods).    This decay occurs because the spacers in a cassette compete to form complexes with Cas proteins; thus, having more distinct spacers effectively decreases the average number of complexes that would be specific to each infection.    Thus, the smaller the cassette size, the more likely that the  $d_c$ specific complexes required for an effective CRISPR response will be produced.   The optimal cassette size is a compromise between this drive towards having less memory, and  the drive to have a defense that spans the pathogenic landscape.

If the detection probability depends sharply on the number of bound complexes with a threshold $d_c$ (Fig.~\ref{fig:numericald3}a), to a first approximation, fewer than $d_c$ complexes bound to a specific spacer are useless, as  detection remains unlikely, and larger than this number is a waste, as  it would not improve detection.  In this case, if the expression of the {Cas} protein was a deterministic process, it would be optimal to have a cassette with $N_p/d_c$ spacers, each of which could be expressed and bind to exactly $d_c$ complexes, predicting that $L = (d_c/N_p)^{-1}$. However,  since  gene expression is  intrinsically stochastic, there would sometimes be more than $d_c$ bound complexes for a given spacer, and sometimes less.   This stochastic spreading  weakens the dependence of the optimal cassette size on the threshold $d_c$, resulting in  the exponent $\beta$ and coefficient $C$ to be slightly less than one in the optimal cassette size scaling $L \sim C (d_c/N_p)^{-\beta}$ (Fig.~\ref{fig:numericald3} insets); {see Methods for a more detailed derivation.}  If the detection threshold is soft,  the CRISPR mechanism effectiveness is less dependent on having at least $d_c$ complexes, specific to a phage. In addition, having a slightly higher  number of complexes than the detection  threshold  can increase the detection probability.  These effects combine to  produce a scaling between the optimal cassette size and the detection efficacy, $L \sim  (d_c/N_p)^{-1}$  (Fig.~\ref{fig:numericald3}b,c).

In summary, our model predicts that a more effective CRISPR mechanism (i.e., having lower detection threshold $d_c$ or a larger number of complexes $N_p$) should be associated with a  greater amount of immune memory (i.e., larger cassette size $L$).

{Our model also provides an  estimate for typical cassette size in bacterial populations countering a diverse set of pathogens (regime of $K \gg N_p$). Assume  that the typical number of {Cas} complexes {is $N_p\sim 1000$},  comparable to the copy number of  other proteins in a bacterial cell \cite{Milo2013,Price2016}, and that {rapid} detection of {an infecting} phage requires a modest number of activated CRISPR-Cas complexes, with a detection threshold in range of  {$d_c\sim 10 - 100$ \cite{Jones1420}}. Our model then  predicts that the optimal CRISPR cassette size should lie in the range of $L\sim 10 -100$, consistent with empirical observations~\cite{Devashish_2015,Martynov2017,grissa2007crisprdb, Barrangou2014,Mangericao2016,Horvath:2008}.% and explaining the small size of the adaptive immune repertoires of bacteria.}

\section{Discussion}
A bacterium's ability to neutralize phage attacks depends on the number of spacers stored in its CRISPR cassette. In laboratory experiments, where viral diversity is limited, a few new acquired spacers per bacterium are enough to stabilize a bacterial population. A Streptococcus thermophilus population defending against continual phage attack acquired at most four new spacers over $\sim 80$ generations, with over $50\%$ of the population having only one new spacer. A different experimental design where individual bacteria with different spacers were mixed found that an overall spacer diversity of $\geq 20$ across the starting bacterial innoculum was enough to stabilize the population. 

 Surprisingly, in natural populations, where viral diversity may be high, cassette lengths are still  small. 
Metagenomic analysis of the human gut microbiome revealed CRISPR cassettes with  twelve spacers on average \cite {Mangericao2016}. A much broader analysis over all sequenced bacteria and archeae found cassette lengths clustered in twenty to forty range~\cite{Grissa:2007}. In agreement with this global analysis, a study targeted at 124 strains of Streptococcus thermophilus revealed an average cassette size of 33~\cite{Horvath:2008}.}   One explanation for these observations may be that a bacterium in the wild encounters only a small set of phages (e.g. due to spatial constraints in habitats).  Alternatively, it may be that the small cassette size results from a functional tradeoff in the bacterial immune system.   Our theory examines both how these considerations -- phage diversity and tradeoffs in the CRISPR mechanism -- affect the optimal amount of immune memory that a bacterium should store.

We identified two qualitatively distinct regimes for the statistics of optimal CRISPR immune memory.   When phage diversity ($K$) is much lower than the number of available Cas proteins ($N_p$),  the CRISPR cassette length $L$ should increase sublinearly in $K/N_p$,  approximately as the square root (Fig.~\ref{fig:numericald2}).   This resembles the previously identified optimal surveillance strategy of ``square-root biased sampling"  to catch rare harmful events while minimizing the wasteful resources spent on profiling  innocents~\cite{Press:2009, Mayer2015b}.

When phage diversity is high ($K \gg N_p$), the number of available Cas proteins $N_p$ limits effective use of memory to mount an immune response.    In this regime, the optimal cassette size increases linearly with $N_p$, $L\sim N_p/d_c$.  Here $d_c$ quantifies the average number of Cas complexes specific to a a given phage that is necessary to mount an effective defense.   Why would the number of Cas proteins be limited in the first place?  Perhaps because there is in general a physiological cost to maintaining high levels of any protein within a cell.  High sustained expression of the Cas system in particular may lead to auto-immune like phenotypes, where a bacterium's lifespan is reduced because of the acquisition of spacers from its own genome~\cite{Jiang2013}.  This may explain why, in some bacteria, the expression of Cas proteins is controlled by a quorum sensing pathway that is triggered when the density of bacteria is high, making them more susceptible to infections  \cite{Patterson2016,Kroghsboe_2012,Hoyland-Kroghsbo2017}.

To understand the tradeoffs that apply to CRISPR cassette size in the wild, we require a better understanding of both  bacterial physiology and  phage diversity in local environments.
 Unfortunately, quantifying phage diversity from metagenomic data is challenging. Phage genomes are constantly changing as they undergo error-prone replications due to sub-optimal use of hijacked bacterial replication machinery. Additionally,  phages often swap   pieces of  DNA  with their  hosts as they are assembled and packaged inside an infected bacterium, leading to drastic changes in their gene synteny. These   large-scale genomic changes challenge the reliability of genome assembly and alignment techniques in quantifying phage  diversity in a given community.  At the same time, the expression levels of  Cas proteins are also not well-quantified across broad bacterial families. However, we can expect that these questions will be addressed by ongoing advances in metagenomics  and in high-throughput expression measurements in bacterial communities.

The authors of~\cite{Martynov2017} proposed a different theory for the small size of CRISPR cassettes.  They suggested that the CRISPR mechanism reads out more recently stored memories first -- i.e., CRISPR is a Last-In-First-Out (LIFO) memory system. In this picture, as old memories become less useful, they are harder to access and will degrade by genetic drift, restricting the size of the ``functional" CRISPR cassette. Our model  proposes  an alternative explanation --- i.e., CRISPR cassettes are small simply because they are adapted for optimizing the immune response in a world where threats are diverse while the CRISPR machinery for detecting and cleaving phage is subject to molecular bottlenecks.   The alternative theories can be experimentally distinguished by better understanding the readout mechanism of the CRISPR cassette.  Of course it could also be that both mechanisms are at play in limiting the utility of deep immune memory.

Vertebrates also possess an adaptive immune system that learns from past infections to defend against future threats.   It has been suggested that pathogen detection in vertebrates is optimized by biasing the immune repertoire towards sensing rare infections with a higher chance than warranted by their frequency~\cite{Mayer:2015}. 
This sort of optimization is unlikely to be  relevant to individual bacteria given the small cassette size and high cost of CRISPR proteins. However,  it is known that across a bacterial population spacer abundances are highly variable but distributed in a stereotypic way~\cite{Bonsma-Fisher:2018}.   It is possible that this stereotyped distribution represents an optimally adaptive immune strategy for the population as a whole.  To study this question,  our framework could be extended to analyze optimal strategies for distributed adaptive immunity in microbial communities.

For both vertebrates and bacteria, a major challenge is to characterize dynamics of the immune system as it chases a diverse pathogenic population that is itself evolving to evade detection by its hosts.
This out-of-equilibrium process can last over an extended evolutionary period~\cite{Nourmohammad:2016}.  Most  models, including the one presented here,  ignore these dynamical  aspects of the co-evolving CRISPR-phage system, which may partly  influence the statistics of stored memory, including the optimal  cassette length. Accounting for these dynamics will shed additional light on the organization of immune memory and the de-novo response to evolving pathogens in organisms with well-adapting immune systems.

\section{Methods}
{\small
\subsection{Probability of successful immune response}
Applying the  assumptions described in the Result section, the complete model (eq.~\ref{eqn:probid}) reduces to,
\begin{equation}
{P_\text{survival} = 1-\left( p_0 + \sum_d p_1(1-\alpha(d)) \, q(d)  \right)}
\label{eq.probNoDet}
\end{equation}
where $p_0$ and $p_1$ are respectively  the probabilities for a bacterium to have $0$ or $1$   spacers specific to the infecting phage, $d$ is the number of specific complexes,
$q(d)$ is the probability of producing $d$ complexes, and $\alpha(d)$ is the probability that a cassette producing $d$ specific complexes recognizes  the phage.  We assume that it is unlikely for a bacterium to carry more than 1 spacer against  a given phage ($p_1\approx 1-p_0$), and  take infections to be random events drawn from a pool of $K$ distinct viruses. Hence, the probability that none of the spacers in a cassette of size $L$  recognizes  an invading phage is $p_0 = (1 - 1/K)^L \approx e^{-L/K}$. The survival probability is 
\begin{equation}\label{eqn:probfinal2}
\begin{aligned}
{P_\text{survival}  =  \left(1-e^{-\frac{\lambda}{\kappa}}\right) \times   \left(1 - \sum_{d < N_p} \big(1-\alpha(d)\big) q(d)\right)} \, ,
\end{aligned}
\end{equation}
where $\kappa=K/N_p$ and $\lambda=L/N_p$ denote a normalized viral diversity and immune  capacity, respectively.
We assume that transcription events occur independently and are equally likely.  Thus,  $q(d)$ in eq.~\ref{eqn:probfinal2} is given  by a binomial density describing the probability of having $d$ complexes specific to a given phage, given $N_p$ Cas proteins and an equal probability of selecting any one of the $L$ spacers to produce each complex.  A successful detection typically requires activation of multiple complexes with a  minimum (critical) number, $d_c$.  Accordingly, we choose the detection probability function  $\alpha(d)$ to be a  threshold function that saturates to  $1$ at   $d\geq d_c$ (SI).

%We assumed that, realistically, transcription events occur independently and with equally likelihood.  Thus $q(d)$ is the product of a binomial factor counting the number of ways of choosing $d$ complexes out of $N_p$,  the probability that the specific spacer will be found in $d$ complexes -- $(1/\lambda N_p)^d$, and the probability that the remaining complexes are bound to other spacers -- $(1 - 1/\lambda N_p)^{N_p - d}$.   Successful recognition typical requires a critical number of complexes $d_c$ {\color{red} Cite}.  Thus, we consider two alternative functional forms for  the probability of recognition $\alpha(d)$ that saturates to $1$ above $d=d_c$. 

\subsection{Optimal cassette size}
The optimal cassette  size relative to the number of Cas proteins, $\lambda = L/N_p$, can be  evaluated by  optimizing the survival probability $\frac{\partial }{\partial \lambda}P_\text{survival}\Big|_{\lambda^*}=0$.
In the biologically realistic regime, where the number of  complexes is both large $N_p\gg 1$, and large compared to the cassette size $ N_p /L\gg 1$, the binomial probability density for the number of specific complexes $d$  in eq.~\ref{eqn:probfinal2} can be approximated by a Gaussian $\mathcal{N}\left(\lambda^{-1} ,{\lambda}^{-1} (1-\frac{\lambda^{-1} }{N_p})\right)\simeq \mathcal{N}\left(\lambda^{-1} ,{\lambda}^{-1}\right)$, up to quantities of order $\mathcal{O}(1/\sqrt{N_p})$. 
In this limit, the optimization criterion gives 
\begin{equation}
%0=1-\kappa (e^{\lambda/\kappa}-1) \sum_{d=0}^{d_c} \mathcal{N}(\lambda^{-1},\lambda^{-1}) \left(\frac{\lambda}{2}-\frac{d^2-1/\lambda^2}{2}  \right)-\sum_{d=0}^{d_c} \mathcal{N}(\lambda^{-1},\lambda^{-1})
0=1-\lambda^* \sum_{d=0}^{d_c} \mathcal{N}\left(\frac{1}{\lambda^*},\frac{1}{\lambda^*}\right) \left(\frac{3}{2\lambda^*}-\frac{d^2-(\lambda^*)^{-2}}{2}  \right) \, ,
\end{equation}
where we assumed a diverse  pool of viruses $\kappa\gg1$ and a sharp  recognition function $\alpha(d)=\theta(d-d_c)$. Approximating the sum in eq.~(\ref{eqn:probfinal2}) with an integral with strict boundaries $[0,d_c]$, we arrive at an equation for the optimal  cassette size  $\lambda^*$, 
\begin{equation}
\begin{aligned}
%0=1-\kappa (e^{\lambda/\kappa}-1) \sum_{d=0}^{d_c} \mathcal{N}(\lambda^{-1},\lambda^{-1}) \left(\frac{\lambda}{2}-\frac{d^2-1/\lambda^2}{2}  \right)-\sum_{d=0}^{d_c} \mathcal{N}(\lambda^{-1},\lambda^{-1})
1=& \frac{-e^{-\frac{1}{2\lambda^*}} +(1+d_c\lambda^*) e^{-\frac{(1-d_c \lambda^*)^2}{2\lambda^*}}}{2\sqrt{2\pi\lambda^*}}\\ 
&+\frac{1}{2}\left(\text{Erf}\left[\frac{1}{\sqrt{2\lambda^*}}\right]+\text{Erf}\left[\frac{d_c\lambda^*-1}{\sqrt{2\lambda^*}}\right]\right) 
%\lambda \sum_{d=0}^{d_c} \mathcal{N}(\lambda^{-1},\lambda^{-1}) \left(\frac{3}{2\lambda}-\frac{d^2-1/\lambda^2}{2}  \right)
\end{aligned}
\end{equation}
In the limit that the cassette size is much smaller than the number of available complexes $L/N_p= \lambda \ll 1$, with  $d_c\lambda$ finite,  the optimal cassette length  scales inversely with the  activation threshold {$L^*\sim \left(\frac{ d_c}{N_p}\right)^{-1}$}.}

\noindent {\bf Acknowledgments: } VB is supported in part by a Simons Foundation grant in Mathematical Modeling for Living Systems (\#400425) for Adaptive Molecular Sensing in the Olfactory and Immune Systems, and by the NSF Center for the Physics of Biological Function  (PHY-1734030).  AN is supported by the DFG grant (SFB1310) for  Predictability in Evolution and the MPRG funding through the Max Planck Society. SG is supported by the NSERC Discovery Grant. VB, SB, and AN thank the Aspen Center for Physics, which is supported by National Science Foundation grant PHY-160761,  for hospitality in the initial stages of this work.\\\\

\newpage{}

\pagenumbering{roman}
\setcounter{page}{1}
\setcounter{equation}{0}
 \renewcommand{\theequation}{S\arabic{equation}}

\onecolumngrid
\noindent {\Large \bf Supplementary Information}\\\\

\subsection*{Optimal cassette size for detection threshold $d_c=1$}

In the case of a switch-like receptor where only one complex is needed to detect the phage, i.e. $\alpha(d)=\theta(d-d_c)$ and $d_c=1$ we get

\begin{equation}\label{eqn:probsimple}
P_\text{survival}= \left(1-e^{-\frac{\lambda}{\kappa}}\right) \left(1-e^{-\frac{{1}}{\lambda}}\right)\, .
\end{equation}
So the condition for the optimal cassette is 
\begin{eqnarray}\label{Sderivative}
 0=\frac{\partial}{\partial\lambda}  P_\text{survival}\Big|_{\lambda^*}=\left(1-e^{-\frac{1}{\lambda^*}}\right) \frac{e^{-\frac{ \lambda^*}{\kappa}}}{\kappa}-e^{-\frac{1}{\lambda^*}} \left(1-e^{-\frac{ \lambda^*}{\kappa}}\right) \frac{1}{\left(\lambda^*\right)^2}.
\end{eqnarray}
 Introducing $f(x)=e^{-x/\kappa} \left(1-e^{-1/x}\right)$ and $a^2 = \left(\lambda^*\right)^2/\kappa$, equation (\ref{Sderivative}) entails,
%$ \frac{1}{\kappa}  f(\lambda^*) = \frac{1}{\left(\lambda^*\right)^2} f\left(\frac{\kappa}{\lambda^*}\right)$.
%Introducing  , equation (\ref{derivative}) follows,
\begin{eqnarray}
 \frac{1}{\kappa}  f(a \sqrt{\kappa}) = \frac{1}{a^2 \kappa} f\left(\frac{\sqrt{\kappa}}{a}\right)
 \label{Smod_der}
\end{eqnarray}
The relationship in equation~\ref{Smod_der} is only satisfied when $a=1$, implying that the ansatz with $a=1$ and $\lambda^*=\sqrt{\kappa}$ is always a solution. Thus, the optimal cassette size follows,
$$L^* ={N_p} \lambda^*= \sqrt{{N_p}K}.$$

As discussed in the main text and shown in Fig.~3, for larger detection thresholds  $d_c\geq2$,  the sublinear relation between the optimal cassette size and the phage diversity is only valid in the regime of low phage diversity $K\ll N_p$. For large phage diversity ($K\gg N_p$) and as long as $d_c\geq2$,  the optimal amount of memory depends only on the CRISPR mechanism  via the response threshold $d_c$, but does independent on the viral heterogeneity (Methods, Fig.~3).

%When $d_c\geq2$, this sort of power-law growth of $L^*$ as a function of $K$ occurs when  $K\ll N_p$, while when $K\gg N_p$ the optimal cassette size is constant and independent of K (see main text).

  \setcounter{figure}{0}
 \renewcommand{\thefigure}{S\arabic{figure}}
    
\begin{figure}[h]
\includegraphics[width=.8\textwidth]{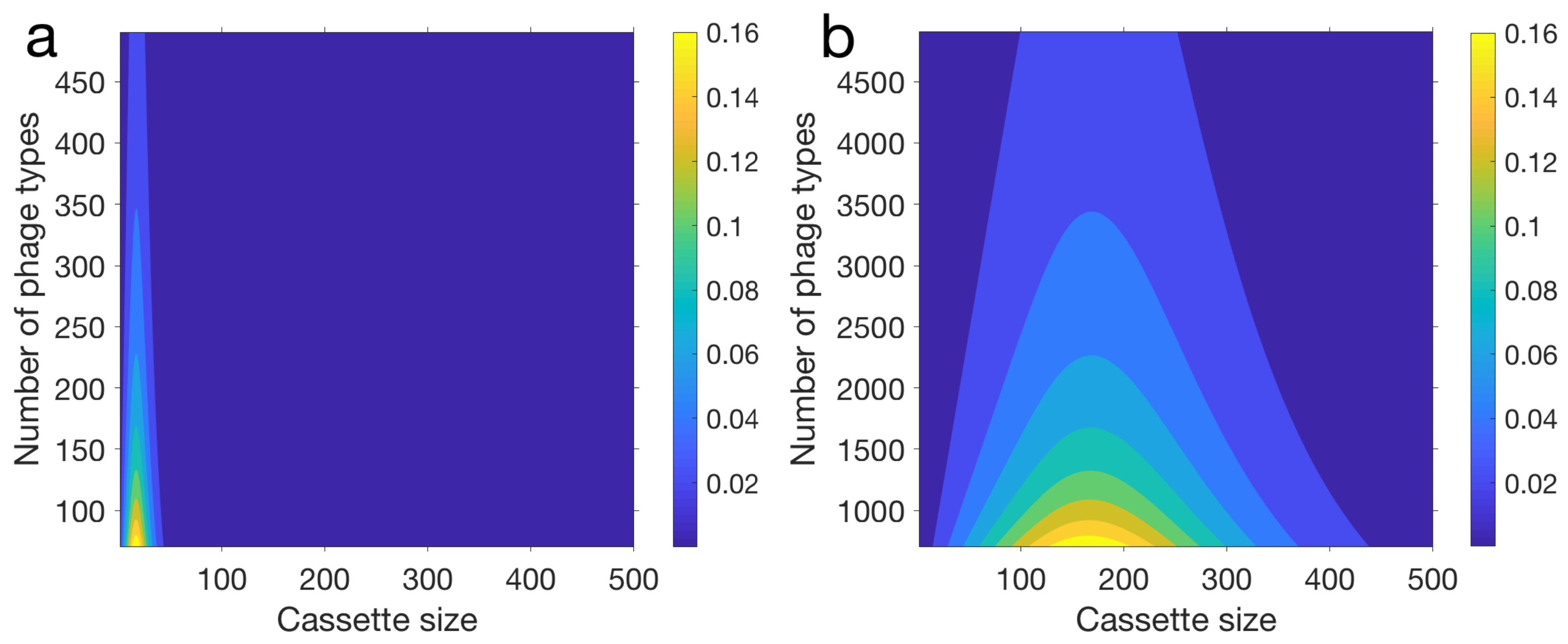}
\includegraphics[width=.8\textwidth]{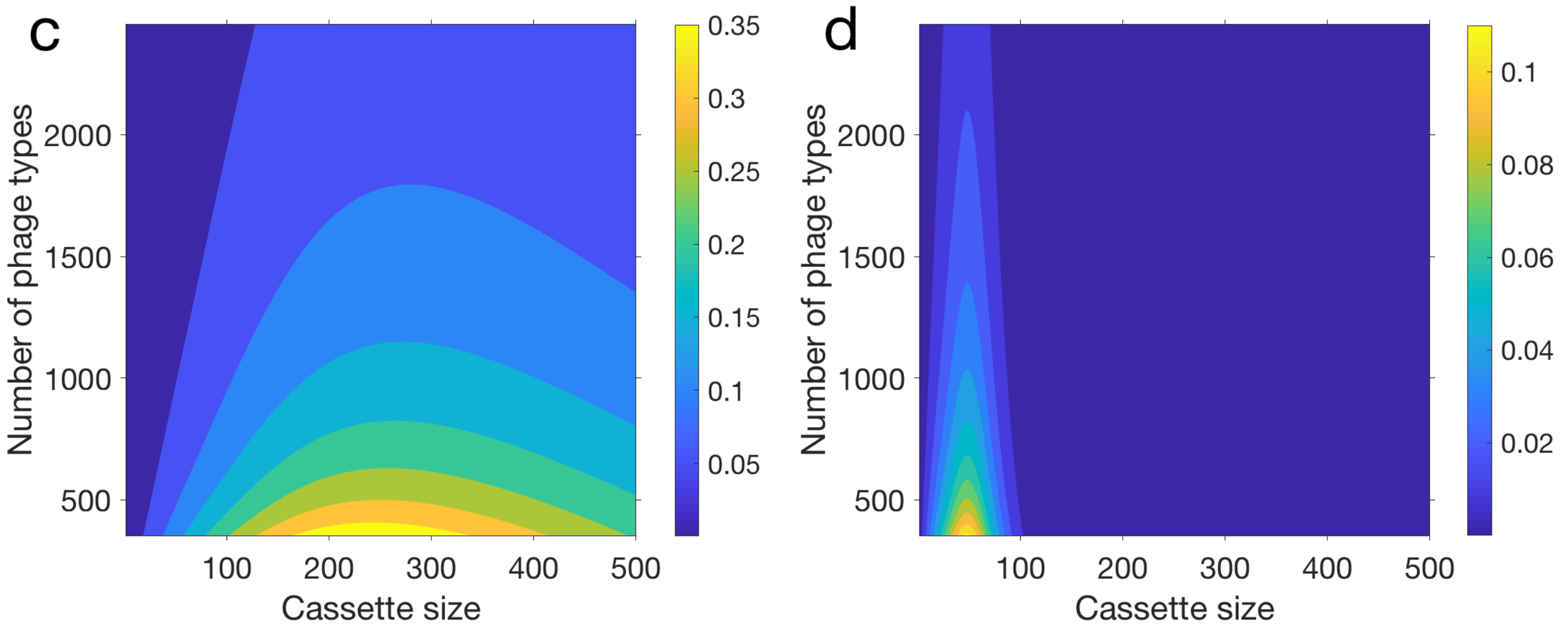}
\includegraphics[width=.8\textwidth]{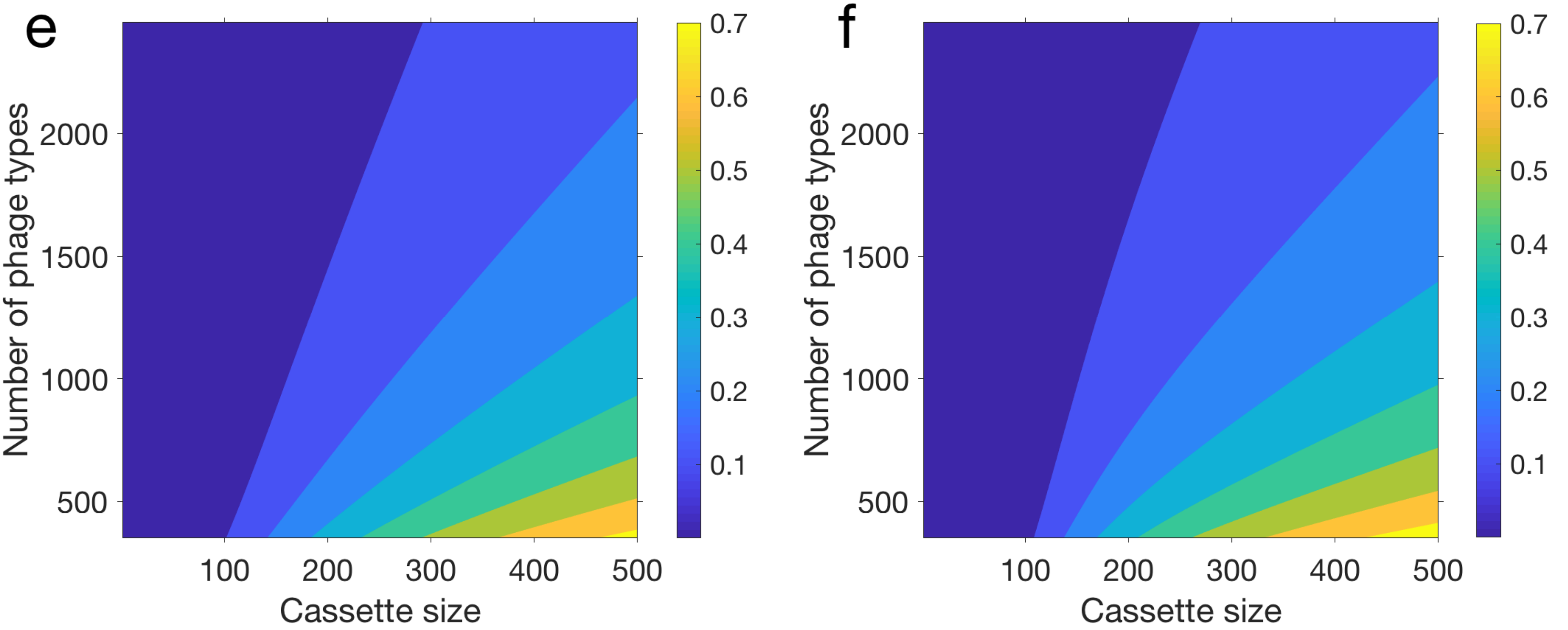}
\caption{{\bf The optimal cassette size depends on  phage diversity and  detection efficacy of CRISPR-Cas.}
The  heatmap shows  the probability of surviving a phage infection  $P_{\text{survival}}$ as a function of the cassette size and the number of phage types $K$.  %$P_{\text{survival}}$ can be interpreted as the fractional population size that will persist after sequential phage attacks if CRISPR is the only defense mechanism. 
Parameters are chosen as follows: {\bf (a,b)} $N_p=200$ and $N_p=2000$, respectively with $\alpha(d)=\theta(d-d_c)$ (step function) with detection threshold $d_c=8$ , {\bf (c,d)} $N_p=1000$ and $\alpha(d)=\theta(d-d_c)$ with $d_c=2$ and $d_c=15$, respectively, {\bf (e,f)} $N_p=1000$ and $\alpha(d) = {d^h \over d^h + d_c^h}$ with $d_c=8$, and $h = 3$ and  $h=10$ respectively. Increasing $N_p$ has a similar effect as decreasing $d_c$. Softening the switch-like behavior by changing the detection probability  $\alpha(d)$ to a Hill-function, we obtain higher values of the optimal cassette size and much higher survival probability.  In all regimes, the survival probability displays a maximum at an intermediate cassette size $L^*$  for all levels of phage diversity.  In other words, for a given environmental condition there exists an optimal cassette size that maximizes the survival probability of a bacterium using  CRISPR-Cas.}
\end{figure}

%\begin{figure}[h]
%\includegraphics[width=0.8\textwidth]{FigS4a.pdf}
%\caption{{\bf Bacterium's survival probability  depends on viral diversity and number of Cas complexes.} The surface plot shows the probability of surviving a phage infection $P_{\text{survival}}$ as a function of the number of Cas proteins and of phage diversity $K$ for the detection probability function $\alpha(d)=\frac{d^3}{d^3+d_c^3}$ and $d_c=10$ evaluated at the optimal cassette size $L^*$. $P_{\text{survival}}$ can be interpreted as the fraction of the  population  that would persist after sequential phage attacks, using CRISPR-Cas as its only defense mechanism.}
%\end{figure}

\end{document}